# Title: Generation of hypercubic cluster states in 1-4 dimensions in a simple optical system

Zhifan Zhou[1], Luís E. E. de Araujo[1,2], Matt Dimario[1], Jie Zhao[1], Jing Su[1], Meng-Chang Wu[1], B. E. Anderson[3], Kevin M. Jones[4], Paul D. Lett[1,5*]

[1]Joint Quantum Institute, National Institute of Standards and Technology and the University of Maryland; College Park, Maryland 20742, USA.

[2]Institute of Physics Gleb Wataghin, University of Campinas (UNICAMP); 13083-859 Campinas, São Paulo, Brazil.

[3]Department of Physics, American University; Washington DC 20016, USA.

[4]Department of Physics, Williams College; Williamstown, Massachusetts 01267, USA.

[5]Quantum Measurement Division, National Institute of Standards and Technology; Gaithersburg, Maryland 20899, USA.

*Corresponding author. Email: lett@umd.edu

**Abstract:**
Entangled graph states can be used for quantum sensing and computing applications. Error correction in measurement-based quantum computing schemes will require the construction of cluster states in at least 3 dimensions. Here we generate 1-, 2-, 3-, and 4-dimensional optical frequency-mode cluster states by sending broadband 2-mode vacuum-squeezed light through an electro-optical modulator (EOM) driven with multiple frequencies. We create the squeezed light using 4-wave mixing in Rb atomic vapor and mix the sideband frequencies (qumodes) using an EOM, as proposed by Zhu et al. (*1*), producing a pattern of entanglement correlations that constitute continuous-variable graph states containing up to several hundred qumodes. We verify the entanglement structure by using homodyne measurements to construct the covariance matrices and evaluate the nullifiers. This technique enables scaling of optical cluster states to multiple dimensions without increasing loss.

Measurement-based quantum computing for continuous variables has been proposed for optical quantum information processing. This one-way quantum computing technique uses cluster graph states, where the initial state and the entanglement required for computation are built into the constructed quantum state, and a series of (possibly conditional) measurements is carried out that encodes the computational program (*2*). The operational quantum bits, or qumodes, in this case are continuous-variable modes of the field. While all of the aspects of this approach to computing have not yet been put together in one system, the type of graph states required for it have been constructed in several different optical systems. Large 1-D optical cluster states (*3-7*) have been created in the time and frequency domains, with up to several million states in the time domain and thousands in the frequency domain. Other approaches in the time-frequency domain and spatial domain have also been developed (*6, 8*). 1-D cluster states are insufficient to implement general gate operations and 2-D cluster states are required for controlled-logic gates and universal quantum information processing. Large 2-D cluster states have also been demonstrated in the time domain (*9, 10*), and recently 2-D cluster states of ≈ 60 frequency modes have been generated using multiple pump frequencies driving 4WM in a silica microresonator (*11*). To implement error-correction codes 3-D cluster states will be required, and perhaps even larger dimensionality will be desired. Unfortunately, the additional optical delay lines required in time-bin-based systems can also introduce additional losses. A method of constructing cluster states in the frequency domain using an electro-optical modulator (EOM) has recently been proposed that is quite general and flexible (*1*), and avoids introducing extra optical elements and loss in scaling to higher dimensions. We implement this idea using a simple optical system based on 4-wave mixing (4WM) in Rb atomic vapor to generate 2-mode squeezing, followed by qumode mixing with EOMs, and demonstrate its ability to generate hypercubic cluster states of up to 4 dimensions.

Graph states are quantum states comprised of individual states or nodes that have generally been entangled in some pattern. A Gaussian cluster state is usually considered to be a class of graph states where the nodes are phase-squeezed states that are sparsely-enough interconnected to allow for measurement-based quantum computing (*12-14*). In conjunction with non-Gaussian resources, fault-tolerant quantum computing is theoretically possible for finitely-squeezed Gaussian states (*15, 16*). A 1-D cluster state can be constructed by mixing light from nearest-neighbor modes together with beamsplitter-like interactions to create entanglement. In the present experiments our qumodes are frequency modes and the beamsplitters are implemented in frequency-space using an EOM. The EOM shifts the frequency of a fraction of the light from a carrier frequency into sideband frequencies, and vice-versa, with a frequency difference, phase, and amplitude that are determined by the radio-frequency electronic drive of the EOM. The relative ease of driving an EOM with multiple frequencies allows us to make a proof-of-principle implementation of this technique for the construction of n-dimensional hypercubic cluster states, as suggested in (*1*).

We have demonstrated a 4WM process in Rb vapor that generates strong 2-mode squeezing into free-space beams (*17-19*) which can easily be passed through free-space EOMs. The experiments, outlined in Fig. 1, are quite straightforward. Because of the large single-pass gain in this system we do not require a cavity to create an optical parametric oscillator (OPO), as discussed in (*1*). Thus, we do not rely on the OPO cavity modes to define our qumodes, but define them after-the-fact, in software, from a continuous spectrum. Because of the limited squeezing bandwidth, our particular 4WM system is not practical for scaling up for computing, but it does allow for the creation of higher-dimensional cluster states with modest numbers (hundreds) of qumodes that can be tested for structure and fidelity.

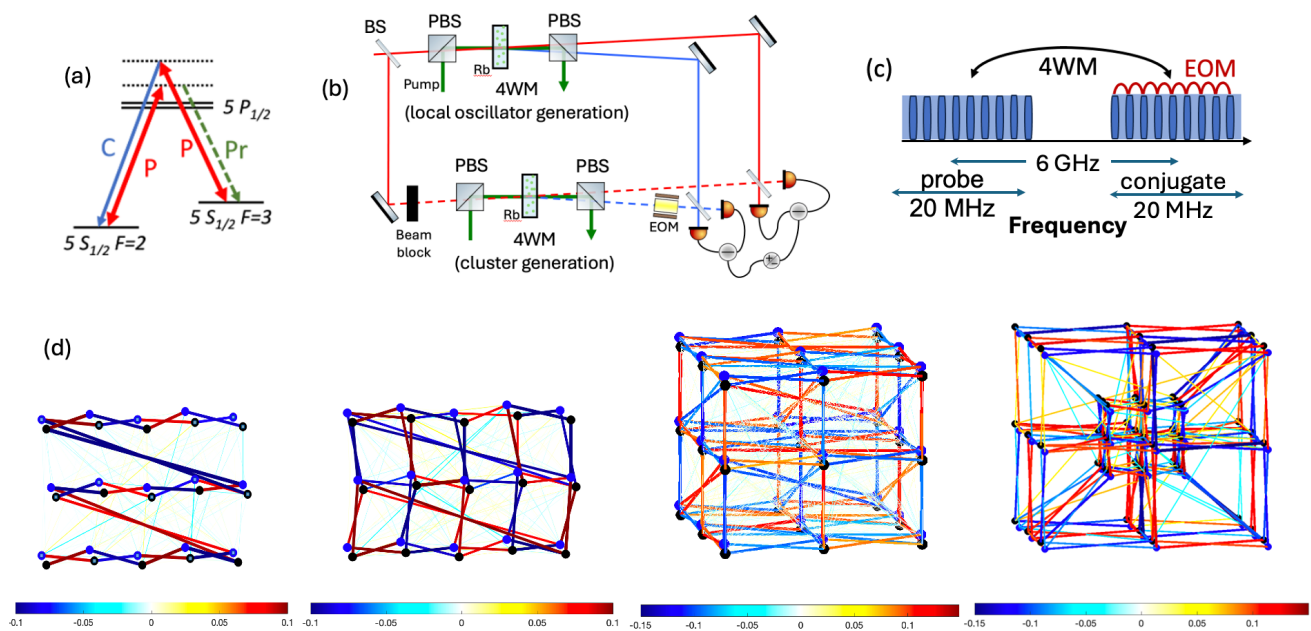

**Fig. 1. Cluster state generation using a four-wave mixing (4WM) interaction in Rb vapor and an electro-optical modulator (EOM).** (a) Four-wave-mixing scheme for generating entangled twin beams in $^{85}$Rb vapor. Two pump photons (P) are converted to a pair of probe (Pr) and conjugate (C) photons. (b) Sketch of the experimental set-up. PBS indicates a polarizing beamsplitter, BS is a non-polarizing beamsplitter. The local oscillators are generated in a separate region of the same vapor cell as the signals, but they are separated here for clarity. The beam block is inserted after alignment of the signals with the local oscillators. Balanced homodyne detection is used at each frequency. EOMs can be placed in each of the twin beams but the experiments can also be performed with a single EOM. (c) Sketch of the frequency-space entanglements created in the experiments. The 4WM interaction entangles frequencies between the twin beams, while the EOM acts as a beamsplitter in frequency space and entangles frequencies within a single beam. (d) Sections of the adjacency matrix graph of experimental 1-, 2-, 3-, and 4-D cluster states created by this method. The blue/black pairs of nodes represent the probe/conjugate qumodes at a given sideband frequency and are strongly entangled by the 4WM process. Nearby frequency modes are mixed by the EOM and the strength of the mixing is given by the covariance values from the adjacency matrix and indicated by the colored connections between the nodes. The case on the left represents a 1 x 15 section of a 100 mode 1-D dual-rail cluster state folded into 3 rows. Smaller, unwanted connections between the nodes are due to noise in the measurements. Traceback connections, for instance, between the lines in the 3x5 section of the 200 mode 2-D case shown, would need to be eliminated for computing purposes; the central 3 x 3 section would be a usable cluster. Sections of the adjacency matrix graph for a 200 mode 3-D case and a 600 mode 4-D case are also shown. (The measurement statistics vary; 12 runs of 10 ms each in 1-, 2-, and 4-D, and 24 runs in 3D. In the 4-D plot the major connections are at a strength of ≈ 0.09, but connections less than 0.05 are not shown for clarity.)

Our 4WM process produces strong 2-mode squeezing over a relatively small bandwidth of around 20 MHz, dependent on the pump intensity (*20*). Detecting these beams using homodyne methods, we have verified that the squeezing at different sideband frequencies is uncorrelated and independent (*21*), leading to a large number of potential independent quantum-correlated qumodes. Recording the time-domain homodyne signals for the two spatial modes generated by the 4WM allows us to employ post-processing of these signals in software, as suggested in (*22-24*). The restricted bandwidth does have the advantage that it allows us to easily capture the entire bandwidth of the quantum-correlated signals digitally and we do not require separate detectors and local oscillators for each qumode, as in previous experiments. We can numerically filter the signals into defined frequency bins and the correlation and entanglement structure as well as the fidelity of the states can be studied off-line. In our experiments we perform homodyne measurements where all the qumodes are detected simultaneously and with the same local oscillator (LO) phase. Such a strategy, where all Gaussian processing after a non-Gaussian operation could be implemented in this way, and the measurements are evaluated in post-processing, is suggested in (*23*).

The use of EOMs in modulating continuous variable entangled beams has recently been explored (*25*). While the proposal in (*1*) suggests using an EOM in each beam of a 2-mode squeezed light source, the nonlocal interference or entanglement properties of the 2-mode squeezed states (*25-29*) allows all of the modulation to take place in a single beam. Putting the same modulation onto both the probe and conjugate beams is equivalent to putting twice the modulation onto one of the beams. The resulting covariance matrices, which completely characterize a Gaussian state, are the same. This simplification allows for even lower loss in the system by only requiring one additional optical element (*25*).

The EOM, when configured as a phase or frequency modulator, does not inherently induce any loss in the system (practically there is $\approx 2\%$ transmission loss) and couples a particular quadrature signal at a carrier frequency into the orthogonal quadrature at the first sideband frequencies, and from the sidebands back into the carrier frequency. This property allows the resulting graph states to be directly transformed into true cluster states by implementing a series of Gaussian local unitary operations, producing a state with a hypercubic adjacency matrix structure and diagonal error matrix (*1, 30*). The signal that is imposed by transferring quantum correlations from, say the X quadrature at a given frequency, to the P quadrature at frequencies separated by the modulation frequency is phase dependent. That is, the XP correlation signal needs to be detected in-phase with the EOM drive frequency to be maximized (see supplemental).

Experimental system
Our system is based on 4WM in an atomic vapor, generating strong 2-mode squeezing (*17, 19*) with a single pump beam detuned just above the atomic D1 resonance frequency. While the generated squeezing is strong, (up to $\approx 10$ dB in bright-beam intensity-difference measurements) matching the local oscillators for homodyne detection has typically limited the observed quadrature squeezing to < 5 dB. A pump beam from a Ti:sapphire laser with approximately 400 mW of power is focused to a $1/e^2$ beam diameter of approximately 650 μm and tuned $\approx 1$ GHz blue of the $^{85}$Rb S - P$_{1/2}$ transition. This beam is sent through an anti-reflection-coated 12 mm long Rb vapor cell at a temperature of about 120 °C. Two-mode vacuum-squeezed states are generated at frequencies separated by the ground state splitting in $^{85}$Rb (approximately +/- 3 GHz from the pump) in two spatial modes, one on either side of the pump, at a small phase-matching angle of about 0.5 degrees to the pump beam. This continuous-variable system is a starting point for generating cluster states. The fact that the 2-mode squeezed states are at different frequencies (about 6 GHz apart) does not alter this construction.

Local oscillators (LOs) for the homodyne measurement are constructed by seeding a similar 4WM process in the same vapor cell with a separate pump beam of similar power (*19, 31*). This creates a pair of independent, bright, phase-locked beams at the two frequencies required, and with appropriate spatial modes (the probe and conjugate frequencies acquire slightly different Kerr lensing in the vapor) (*19*). One or both of the vacuum-squeezed signal beams are passed through EOMs. The appropriate signal and local oscillator beams are then mixed on 50/50 beamsplitters and sent to photodetectors to implement a balanced homodyne detection scheme (see Fig. 1). The homodyne detection resolves just one frequency range (probe or conjugate) in each beam and selects a small number of correlated spatial modes in each beam as well. We thus detect a range of frequencies in each beam that constitutes many entangled pairs.

The relative phase of the LOs is locked using a noise-locking technique (*25*) to fix the measurement to either the maximum squeezing or the maximum noise phase by feeding back to mirrors in the LO paths, allowing us to measure XX or PP quadrature correlations between the beams. (The quadrature noise is completely symmetric with phase in this 4WM system and the absolute phase is allowed to slowly drift in the experiments.) In order to measure correlations between the X quadrature of one beam and the P quadrature of the other, we use a separate noise-locking system. This locking system allows us to lock the relative phase of the two beams at an arbitrary ratio of the measured noises from the two homodyne detectors, and thus to a point half-way between the squeezing maxima and minima (*25*). The signals from the balanced detectors are amplified, digitized, and individually recorded.

We first examine the frequency-dependence of the 2-mode squeezing or Einstein-Podolsky-Rosen (EPR) correlations that are directly produced by the 4WM process with no drive to the EOMs. The squeezing spectra (red curves) in Fig. 2 are taken with the EOMs in place but turned off. Turning on an EOM removes light from a given frequency bin and mixes light from other frequencies into this bin; the uncorrelated light from other frequencies, while generating entanglement, adds noise to the simple 2-mode squeezing variance. A demonstration of this is seen by measuring the 2-mode squeezing at a given frequency, turning on the EOM with a relatively large drive voltage and seeing that the squeezing is reduced or even converted to excess noise. This 2-mode squeezing measurement with the EOM on is shown in Fig. 2 as the blue curve. The horizontal axis in these plots is labelled by mode number. In this case the frequency modes are 180 kHz wide and separated by 200 kHz (20 kHz is left between the mode bins; see supplemental). The 100 modes shown, spaced by 200 kHz, span sideband frequencies of 0 MHz to 20 MHz.

Results
The nullifiers, (discussed in more detail in the supplemental material) are operators on the ideal entangled cluster states that return the eigenvalue of 0, or in case of imperfect squeezing, the squeezing level. They can be expressed in terms of the elements of the covariance matrix, as discussed in (1, 32). They are variance-based entanglement witnesses that can be used to verify the structure and fidelity of a cluster state. Since in this case the mixing of modes is relatively small, and because the initial squeezing varies with frequency, it is convenient to order the modes based on the sideband frequency, as with the modes of the 2-mode squeezing spectrum. As one can see in Fig. 2, the 2-mode squeezing is partly destroyed by turning on the EOM. The entanglement is preserved, but it is now spread amongst more than the original two modes. Calculating the appropriate sum of covariance terms, we arrive at the nullifier values plotted in Fig. 2. There are two calculations of the nullifiers displayed (see the supplemental material); one from a straightforward covariance extraction from the data, and a second using a software lock-in technique that compensates for any phase shifts between the drive frequencies. The fact that these two agree means that we have appropriately compensated the initial drive waveform for the phase shifts that otherwise arise in the high-voltage amplifier. The nullifier variances for each quadrature return the original squeezing levels, indicating that the entanglement is preserved after the mode-mixing. A section of the adjacency matrix for this state is shown in Fig. 1(d).

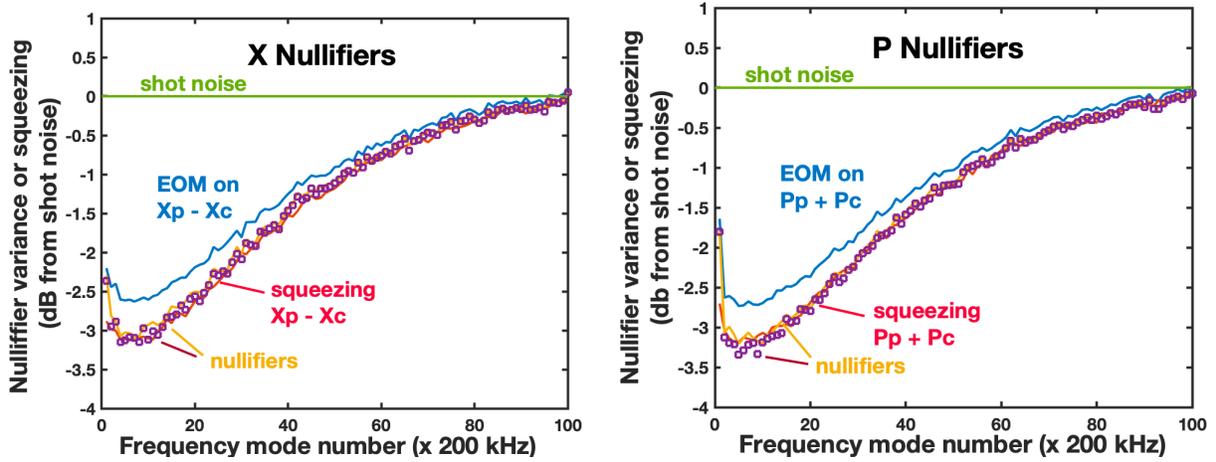

**Fig. 2. Measurements of 2-mode squeezing and nullifier variances as a function of mode number for a 1-D cluster state.** The bin centers are separated by 200 kHz on the horizontal axis, spanning 10 MHz in frequency. The red curves are the 2-mode squeezing variances, measured with the EOM off ($X_p - X_c$ on the left and $P_p + P_c$ on the right) for the particular frequency modes. The blue curves represent the 2-mode squeezing variances for the corresponding mixed modes, measured with the EOM on with a modulation index of 0.18. The circles and the gold lines indicate the appropriate X- and P-nullifier variances calculated for these modes; the lines are from a simple covariance calculation and the circles are obtained using a numerical lock-in detection calculation of the XP correlations. Data is averaged over 12 runs of 10 ms each.

The covariance matrix completely describes the quantum statistical properties of a Gaussian state. The structure of the correlations imposed by the EOM drive frequencies will appear in the XP blocks of the covariance matrix. We choose the frequency width of the qumodes to be spaced by the lowest of the drive frequencies. To construct hypercubic correlation structures the modulation frequencies should be multiples of each other. The modulation index of a sinusoidal modulation on the EOM must be kept small for the resultant graph state to be useful for measurement-based computing. This condition, detailed in (*1*), restricts the entanglement to nearest-neighbor-type connections at a given drive frequency. A 3-D case with modulation frequencies of 100 kHz, 300 kHz, and 900 kHz added together is given in Fig. 3. The entire covariance matrix and an XP block of the covariance matrix are shown. The structure in the XpPc block of the covariance matrix (where p and c denote the probe and conjugate modes) shows 6 off-diagonal terms reflecting the fact that three driving frequencies are applied to the EOM. The nullifier measurements for this 3-D cluster state are shown in Fig. 4, again reproducing the initial squeezing levels, and a section of the adjacency matrix is shown in Fig. 1(d). The modulation index for each individual frequency in this case is 0.18. The nullifiers for a 4-dimensional cluster state driven with sine wave frequencies of 33 kHz, 99 kHz, 297 kHz, and 891 kHz and a modulation index at each frequency of 0.18, are shown in Fig. S4.

As shown, we have generated up to 4-dimensional hypercubic cluster states in this fashion, with varying numbers of qumodes. With low-dimensional clusters we operate with about 100 modes spaced by 200 kHz, for which we average 12 traces, each 10 ms long. For 3-D and 4-D clusters we have used frequency bins as small as 33 kHz (≈ 660 modes over 20 MHz) and average 12 or 24 data traces for reasonable statistics within a frequency bin. Our measured 2-mode squeezing levels are only approximately -3 dB over a small part of the frequency range used, and would need to be improved to beyond -3 dB to claim complete inseparability based on the van Loock-Furusawa criterion, as in (*10*). The graph states directly generated here can be transformed by Gaussian local unitary transformations into true cluster states (*1*) (also see supplemental material).

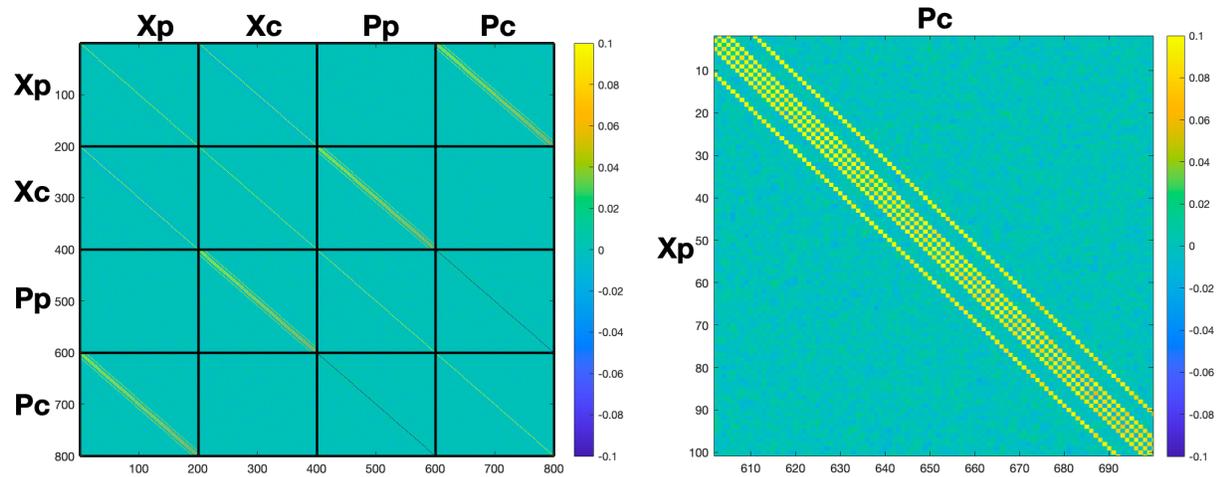

**Figure 3. Covariance matrix for the 3-D cluster state in Fig. 4.** The drive frequencies are 100 kHz, 300 kHz, and 900 kHz applied to a single EOM. The modulation index is m = 0.18 for each frequency. The data is averaged over 24 traces of 10 ms each. The axes are labelled by mode number and the 200 modes in each block span 20 MHz. On the left is the full covariance matrix on an expanded color scale and on the right is a block showing XpPc correlations for 100 of the modes in that block. (In the full covariance matrix the diagonal elements have a magnitude as large as ≈1.5. Same-beam correlations, such as XpPp and XcPc, are not measured and these blocks are set to zero.)

We estimate the error vector (Eq. (36) from Ref. (*1*)) and find that, for our 1-D, 2-D, and 3-D cases we meet the criteria that the elements of this vector be <<1. This should ensure that an ideal cluster state is approached as the squeezing increases. These elements are ≈ 0.2 for the 4-D case presented here, and thus for a useful cluster state we would need to reduce the modulation index or the measurement error to meet the criteria. In any case, to be useful for computing the measured squeezing would need to be increased as well. While not allowing for general computations, this simultaneous-measurement / post-processing implementation would allow the completion of Gaussian operations with a quantum depth of one at the end of a more complicated computation. More complex non-Gaussian operations would require a non-Gaussian input state (*23*). Alternatively, with larger mode spacings individual qumodes could be addressed with non-Gaussian operations.

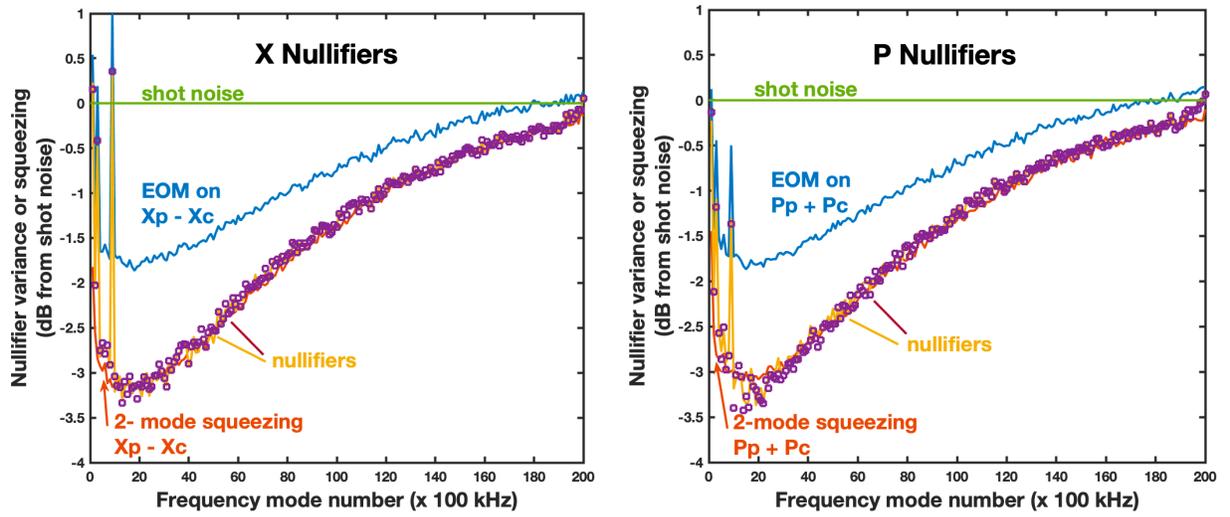

**Figure 4. Two-mode squeezing and nullifier variances for a case of a 3-D cluster state.** The mode centers are separated by 100 kHz on the horizontal axis, spanning 20 MHz in frequency. The drive frequencies are 100 kHz, 300 kHz, and 900 kHz, applied to a single EOM. The modulation index is m = 0.18 for each frequency. The plots show the amplitude-difference squeezing and X nullifiers (left), and phase-sum and P-nullifiers (right). The red curves are the 2-mode squeezing variances with the EOM off for the particular frequency modes. The blue curves represent the 2-mode squeezing variances for these modes with the EOM on. The circles and the gold lines indicate the appropriate X- and P-nullifier variances calculated for these modes; the lines are from a simple covariance calculation and the circles are obtained using a numerical lock-in detection calculation of the XP correlations. Large spikes in the nullifiers appear at the frequency multiples (3, 9) of the lowest mode index, where the excess noise in this bin gets mixed into the calculated nullifier variance.

While incorporating the EOM into one of the 2-mode squeezed beams and modulating creates an entangled cluster state, we can accomplish equivalent measurements by putting the EOM into the LO beam(s) (*25*). This has the effect of measuring the appropriate superpositions of two-mode squeezed states so that the measurements would reproduce the covariance matrix - effectively a form of compressed sensing. In fact, we could dispense with the EOM(s) altogether and simply measure the 2-mode squeezing, digitize the results, and create the appropriate superpositions in software, as suggested in (*22-24*), but these approaches would not create the desired quantum states.

Conclusion

These experiments constitute a proof-of-principle demonstration of the novel cluster state generation protocol proposed in Ref. (*1*). The demonstration of 3-dimensional cluster states, necessary for the implementation of many kinds of error correction is important. The use of volume error-correcting codes, instead of surface codes, could perhaps take advantage of 4- and higher-dimensional cluster states.

The optical and electronic system that we use is not challenging in the sense that the laser is free-running and there are no phase-locked loops required for the laser system; the only locking is for the homodyne phases, making it experimentally quite accessible. In our current system the 4WM bandwidth is limited and thus the measurements will necessarily be slow to resolve the qumodes, and access to the individual qumodes is also experimentally limited. The approach demonstrated here, however, could potentially be applied with other 4WM or parametric downconversion processes with larger bandwidths, which would allow faster measurements and more qumodes, as well as the ability to disperse the qumodes sufficiently for individual addressing. Using the originally-envisioned OPOs (*1*) to generate the initial squeezing would also allow for more, and more widely-separated, qumodes.

We have used free-space EOMs with very low insertion loss placed into the squeezed light beams. As such it is a very compact and scalable approach to generating higher-dimensional cluster states. Travelling-wave EOMs and periodically-poled, waveguide-based EOMs allow modulations at GHz frequencies but may also have to contend with larger insertion losses. Depending on the particular application (sensors vs. computing), placing the EOM into the local oscillator for the measurement of appropriate mode superpositions might be an acceptable approach, removing this concern. Using such EOMs with other parametric processes with larger bandwidths, would also provide more flexibility. These changes could make the EOM a particularly valuable tool for both quantum information processing and for sensing arrays.

**Acknowledgments:**
We would like to thank Alessandro Restelli for electronics help, and to acknowledge helpful conversations and advice from Olivier Pfister.

**Funding:**

Air Force Office of Scientific Research grant FA9550- 16-1-0423.

L. E. E. de Araujo acknowledges the financial support of grant #2019/24743-9, São Paulo Research Foundation (FAPESP).

Part of this research was performed while Matthew DiMario held a National Research Council Research Associateship at NIST.

**Supplemental Materials and Methods**
<u>Construction and independence of frequency modes</u>

      We form our frequency qumodes conceptually as frequency bins that are then explicitly defined in software. Other systems, such as optical parametric oscillators (OPOs) based on parametric downconversion (*5, 7*) produce squeezing in successive cavity modes over a broad spectrum, providing a natural definition and separation of the qumodes in those systems. In Refs. (*3, 4, 9, 10*) the continuous-wave generation of squeezed light is broken into imaginary time bins; that is, one imagines the continuous-wave squeezed light that is produced from an OPO as a series of pulses or temporal modes on which successive measurements can be made. These time bins or pulses are then mixed and manipulated to construct cluster states. Here we make a similar construction, although in frequency space.

The Rb 4-wave mixing (4WM) system generates multi-spatial-mode, as well as multi-frequency-mode, 2-mode squeezed states. Here we restrict our consideration of the spatial mode structure of the 2-mode squeezed states (probe and conjugate beams in the language of 4WM) and lump the multiple spatial modes that we detect into two composite "probe" and "conjugate" beams. Using spatial modes from this system to construct cluster states has also been suggested (*33*). Generating the squeezing in a cavity or spatially-filtering the modes of interest would make the signal modes well-defined and surrounded by vacuum modes, and should improve the ability to measure larger squeezing (*34*), but including a cavity entails other experimental complications and instabilities.

Our frequency qumodes are defined with a gap between them. For the measurements here the frequency bins span 90 % of the frequency space. For a 200 kHz bin spacing, for example, we create bins of 180 kHz width, leaving a buffer of 20 kHz between the bins. For the square-edged bins used here this leaves 20 kHz worth of unused data between the bins. Square-edged filter functions constitute non-causal filters. For causal filter functions the filter edges are not sharp in frequency space, and successive bins would overlap to some degree if they were not spaced with such a gap. The mode axes in Figs. 2 - 4 in the main text and Figs. S1 - S5 are labelled by mode number. For the horizontal axes in Figs. 2 and S1 the frequency modes are 180 kHz wide with mode centers starting at 100 kHz and separated by 200 kHz. For the 2D and 3D cases the mode spacing is 100 kHz, starting at 50 kHz, with 90 kHz wide bins, and for 4D the mode spacing is 33 kHz, with 30 kHz wide bins.

The 2-mode squeezing from the 4WM process used here is generated over a bandwidth of approximately 20 MHz. While the generated squeezing actually extends to very low frequencies (see (*21*)) rf splitters in the current measurement chain rolls-off the low-frequency response at about 200 kHz. At higher frequencies the squeezing tracks the 4WM gain and gradually approaches the shot noise level at sideband frequencies around 20 MHz. We have shown that the individual sideband frequencies function as independent sources of 2-mode squeezing over this bandwidth (*21*). We have verified that, down to frequency bin widths of 5 Hz, neighboring bins are uncorrelated and are thus able to be used as the basis of qumodes in this construction of cluster states.

One can see this by correlating the noise in one frequency bin of the probe with different overlapping and non-overlapping frequency bins in the conjugate beam. We first lock the homodyne detectors to fix the LO phases and determine the probe and conjugate quadratures to measure. With the measured signals we implement a digital filter and decide on frequency bin widths and central frequencies to correlate. We then calculate the covariances for the different bins and bin overlaps. We have verified the independence of non-overlapping bins for sharp-edged (non-causal) filters. To within the noise, the 10 % gap between the bins provides the same mode independence for modes determined by reasonably-chosen causal Chebyshev, Butterworth, or elliptic-type filters.

The squeezing/entanglement bandwidth of the present system is limited. Consequently, the frequency bins will be narrow, requiring relatively long measurement times in this implementation. Other systems of 4WM in fibers or parametric downconversion (*5, 7*) can be much more broadband. Such squeezing sources may not have as high of a gain as is present here and may require cavity enhancement in an OPO. Also, the fact that we digitize the signals for all of the qumodes at once in this system, while an advantage for data-taking, limits our ability to introduce non-Gaussian operations (for example, photon counting) on individual qumodes. A broader gain-bandwidth, coupled with electro-optical modulators (EOMs) with GHz modulation bandwidths, could allow for more and wider modes and better mode separation, allowing this technique to applied in a more scalable way.

Measurement details/signal processing

Post processing of the digitized signals allows us to choose various frequency qumodes to correlate and compare, but the signals first need to be appropriately normalized and time-shifted. To accommodate somewhat different electronic gains and local oscillator (LO) powers between the signal paths we can normalize the two balanced detector signals to one another to maximize the squeezing. In addition, the 4WM process creates a group velocity difference between the probe and conjugate beams in the vapor cell (*19, 35, 36*). This delay between the probe and conjugate modes leads to oscillations in the measured squeezing spectrum. For short delays, as is the case here, the long-period oscillation in the spectrum makes the squeezing bandwidth appear to be smaller. This group velocity delay difference (approximately 10 ns, depending somewhat on pump detuning) can be taken out or measured in software by delaying one of the signals, again optimizing the calculated squeezing spectrum, as shown in Ref. (*21*). Once the delay is known it can be compensated by inserting a cable delay in one signal arm prior to the digitization. This is what is done in the present experiments: an approximately 10.4 ns delay is implemented after the detector in the conjugate arm. At the optimal delay the squeezing spectrum rises slowly and monotonically from low frequencies to approach the shot noise level.

This normalized and appropriately-delayed data can then be filtered and the correlations, 2-mode squeezing, or excess noise as functions of the various qumode frequencies can be calculated. We introduce EOMs into the beam(s) (indicated in Fig.1 of the main text) to entangle the frequency modes.

At a small modulation index the EOMs will mix the light from one frequency mode into frequency modes above and below it, spaced by the modulation frequency. A phase modulator driven sinusoidally at frequency $\Omega$, operating on a monochromatic carrier field of frequency $\omega$ mixes light into harmonic sidebands at frequencies $\omega_n = \omega + n\Omega$, where n is an integer, with amplitudes given by Bessel functions of the first kind $J_n(m)$:

$$e^{i[\omega t + m \sin(\Omega t + \phi)]} = e^{i\omega t} \sum_{n=-\infty}^{\infty} J_n(m) e^{in(\Omega t + \phi)} \qquad (S1)$$

where *m* is the modulation index and $\phi$ is the phase of the EOM driving field. This mixes light from one frequency to another, but also, by virtue of the phase shifts in a frequency modulator, it moves light (and quantum correlations) from one quadrature to the other. Thus, some of the correlations between fields in the P quadrature get moved to the X quadrature at the new frequency. And, of course, some of the light and correlations from the X quadrature of that frequency get moved into the (minus) P quadrature of the original frequency as well. This action of moving the light between different frequencies is sensitive to the phase of the drive on the EOM. Thus, the distribution of the quantum correlations in the signals must be detected with this in mind – that is, for a given measurement interval the structure of the XP quadrant of the covariance matrix of the quantum state is sensitive to the EOM drive phase or, equivalently, the detection phase. If covariances are calculated directly from the digitized data, then the data window also should be a whole number of periods of the EOM drive frequency. Alternatively, a lock-in detection method can be employed, multiplying the data by a sine wave at the EOM drive frequency, with the appropriate phase.

As suggested by (*26*) and in experiments looking at photon correlations (*27*), the frequency mixing and quantum correlation movement introduced by an EOM onto a two-mode squeezed state is a nonlocal property of the two-mode system. Indeed, we have explored the placement of the EOMs in the experimental configuration (*25*) and have shown that the two-mode squeezed state can be modulated equally well with in-phase modulation indices of value *m* on EOMs in each beam, or with a modulation index of value 2*m* on a single EOM, and that the EOM can be in either beam. Since the EOM is a linear device it also does not matter if multiple frequencies are imposed on a single EOM. The quantum states involved are Gaussian states and are completely characterized by their covariance matrices. It is straightforward but tedious to show that the states created in the different EOM configurations described above have identical covariance matrices (except for modes near the edge of the spectrum, which mix-in light from frequencies not being considered). This makes driving with two EOMs (one in each beam) unnecessary. Nonetheless, if one wanted to drive with EOMs in each beam and at high frequencies, one should take into account the optical group velocity delay that appears between the beams before the detectors.

The EOMs were chosen to produce minimal wavefront distortion with modulation, so as not to alter the homodyne detection efficiency with modulation. Both probe and conjugate colors are present in each beam, as the 4WM process generates vacuum squeezing in a symmetric cone around the pump beam and in a number of spatial modes across and around this cone. Even the light passing through the EOM(s) to the detectors contains both the probe and the conjugate frequencies in multiple spatial modes and the local oscillators select the detected modes. Our EOMs have a half-wave voltage of 260 V, and with a typical drive voltage amplitude of 30 V at an individual frequency the modulation index is 30 π/520, or approximately m = 0.18 (≈ 0.06 π). This is in the regime discussed in (*1*), and results in only nearest-neighbor entanglement connections being important. Thus, it can be used to construct cluster states suitable for quantum computing operations.

The available voltage amplifiers in our experiments had a limited bandwidth, leading to varying phase delays for different modulation frequencies in the amplifier. While XX and PP quadrature correlations are not phase-sensitive, the XP correlation signals are sensitive to the EOM drive phase, and it is necessary to compensate for the differing phase shifts (and potentially the amplitude response of the amplifier as well) when driving with several frequencies at once. We drive the EOM with a signal from an arbitrary waveform generator programmed to output the sum of the frequencies that we wish to use. We measure the phase shifts and amplitudes through the amplifier and adjust the input waveform so that the drive signal at the EOM is composed of a sum of sine waves at the desired frequencies that are all in-phase and equal amplitude at the EOM at the start of the drive. (Lock-in detection at the different frequencies could extract the appropriate signals at different phases for each frequency, but this could complicate any computing operations.) The higher frequencies are all multiples of the lower drive frequencies in order to form square lattices. The choice of frequencies can be important and is discussed further in relation to the Gaussian local unitary transformations to generate cluster states below.

Other modulations, with multiple sidebands and stronger connections between more qumodes, could be useful for producing entangled sensor networks but such connections would interfere with the current approach to measurement-based computing. Instead of just summing different-frequency sine waves to create cubic graph structures, as is done here, both the phase and amplitude of the sidebands can be designed more intricately using specialized waveforms, as in Ref. (*37*). Graph states such as those constructed here could also be employed in entangled sensor networks (*38-40*) or in Gaussian Boson sampling applications where 3-D graph-type states have been generated for this purpose (*41*).

While correlating the appropriate probe-conjugate quadrature combination (evaluating say, the amplitude-difference $X_p - X_c$, where the subscripts p and c refer to the probe and conjugate beams) will result in squeezing in the best case, correlating with the "wrong" combination (180 degrees out of phase, or amplitude-sum $X_p + X_c$) will give excess noise. Similarly, the variance of the phase-quadrature sum ($P_p + P_c$) will show squeezing. We can also lock the signals to some arbitrary pair of quadrature phases, such as at 90 degrees where we can look at the correlations between $X_p$ and $P_c$. The lock circuits for the XX/PP and XP locking are given in the supplementary material to (*25*).

Driving with a single frequency on an EOM placed in, for example, the conjugate beam, light will be moved from the $X_c$ quadrature to the $P_c$ quadrature in neighboring frequency modes. Correlations will then appear between an $X_p$ mode and the connected $P_c$ modes. Similarly, light will be shifted from $P_c$ into neighboring $X_c$ modes and correlations will appear between these modes and the central $P_p$ mode. The digitized homodyne signals, $X_p$ and $P_c$ for instance, are multiplied together and averaged over the measurement time, integrating over a whole number of periods for each EOM drive frequency, all starting with the same phase. The signals can be extracted with a varying starting time (EOM drive phase) and duration, optimizing the phase, and making the duration a whole number of periods. The signals can also be extracted from the measured data with a lock-in type of detection in software (multiplying the signals by a set of sine waves at the desired frequencies that are in-phase with the signals). In this way the phase-sensitive XP correlation signals can be extracted. The lock-in detection, implemented in software, allows us to extract the optimum signal phase at each frequency and verify that we have optimized this in the hardware as well. Both methods of extracting the XP correlations lead to the (same) optimized phase-sensitive signal, as indicated by the two nullifier calculations shown on Figs. 2, 4, S1, and S4.

The electronic signal amplifier that we use has 36 dB gain and a 39 MHz bandwidth. We typically digitize $10^6$ points, spaced every 10 ns, with an 8 bit digitizer, for a 10 ms long trace. We take 3 to 6 such traces and extract the frequency covariances and then average them. We digitize signals from both the probe and conjugate homodyne detectors, with each detector locked on either the X or the P quadrature. Successive measurements of the XX, PP and XP correlations allow us to fill out the interesting portions of the covariance matrices for the states. (We do not measure correlations between different frequency components of different quadratures of the same beams, for instance correlations between $Xp(\omega)$ and $Pp(\omega')$, since no interaction in the present experiments should create these correlations and they do not appear in any nullifier calculation.)

With just the XX and PP measurements we can calculate the 2-mode squeezing levels. The XP measurements capture the correlations induced by the mixing of the light between the modes caused by the EOM. We measure the shot noise level (vacuum; with blocked signal ports) with just the LOs present. We also measure the starting squeezing level (squeezed light into the signal ports, but with the EOM turned off). Finally, we take measurements with the EOM(s) turned on.

Data manipulation
The data structure consists of 10 ms of data from one probe and one conjugate quadrature measurement, taken as separate XX, PP, and XP measurements; the XP measurement is duplicated for use in the PX quadrant of the covariance matrix. Thus, there are effectively 6 (or 8 if you count PX separately) data traces in a run, and separate runs for shot noise, EOM-off, and EOM-on. Typically, we take 3 to 12 data runs of 10 ms and average (that is, we average the covariance matrices after they are derived from each run).

An individual X or P homodyne measurement is then filtered into frequency modes, as discussed below. A squeezing spectrum can be obtained from either the XX or PP data using several approaches; we can calculate the autocorrelation and Fourier transform it, we can calculate the power spectrum directly, or we can bin the data into frequency bins and calculate the variances, and compare these to similar measurements for shot noise. These are all equivalent calculations.

We have used a sharp frequency filter to define our frequency bins. We create a vector of center frequencies for the filtering, and then set the mode width and bandpass width. These are chosen so that the filtered frequency "modes" do not fill the space - that is, so that there is some space between the frequency mode "edges." The modes are typically chosen to take up 90 % of the frequency space and there is some 5 % of the range at the top and bottom of the frequency bin that is not used. With a causal filter this is because the edges of the filter are not sharp and this leaves some space for the edges of the filters, so that neighboring bins don't overlap. With the "sharp" filter this is not really a problem. At the current level of analysis the use of a causal versus the sharp-edged filter has not produced any noticeable differences.

Given a vector of the center frequencies of the frequency modes, we numerically Fourier transform (FFT) the data, and then mask the data to isolate and filter the frequencies into the various modes. To calculate the covariances between the different frequencies the data is shifted to all be effectively at some common demodulation frequency. Since the Fourier transformation gives both positive and negative frequencies, the demodulation needs to be done carefully. Then the data is inverse-Fourier-transformed back into the time domain, and the covariances between the different frequency bins can be calculated.

Alternately, for the phase-sensitive XP correlations, we can perform a "lock-in" calculation, which is to take the probe and conjugate time traces, FFT them into the frequency domain, filter these two signals into frequency bins (without demodulating them to a common frequency), and inverse-FFT them back into a set of time traces for the frequency bins. One can then multiply together the time traces and a sine wave at the EOM frequency: $X(f_1)*P(f_2)*\sin(\Omega*t)$, where $X(f_1)$ and $P(f_2)$ are the time traces filtered at frequencies $f_1$ and $f_2$, and $\Omega$ is the EOM drive frequency. Integrating or averaging over this product yields the lock-in signal. Doing this with both a sine and a cosine function allows one to extract the amplitude and the phase for the maximum signal. One can then measure different signal (EOM drive) frequencies at different phases in the data or, preferably, adjust the input drive phases to obtain the maximum signal at all frequencies. The two methods of extracting the phase-sensitive XP correlations yield exactly the same results.

Nullifiers
The nullifiers are variance-based entanglement witnesses that can be used to verify the structure and fidelity of a cluster state. The nullifier operators acting on a state return the eigenvalue of 0 (or the squeezing value for finite squeezing levels) and are given by: $P_j - \sum_{k \in N_j} Q_k$, where $j$ and $k$ are mode indices, $P_j$, and $Q_j$, are general momentum and position operators for mode $j$, and $N_j$ is the set of such modes that make up the graph. For a 2-mode squeezed state one can view the simple 2-mode squeezing variances (amplitude-difference and phase-sum for the probe and conjugate modes, or $Xp - Xc$ and $Pp + Pc$) as being analogous to a nullifier variance for a 0-dimensional graph state. Further interconnecting such 2-mode squeezed states creates higher-dimensional graph states with more complex nullifier expressions (linear combinations of the 2-mode nullifiers) which should also reproduce the 2-mode squeezing variances.

2-D and 4-D results
An example of the nullifiers obtained for a 2-D graph state with 100 kHz + 500 kHz modulation waveforms added together is given in Fig. S1. Comparing to Fig. 2 (the 1-D result), one can see that the EPR nullifier variances (2-mode squeezing variances) measured on the new state are further away from the true nullifiers (or the original squeezing measurements) for this state. Generally, this happens as more light is moved around by the EOM as more modulation frequencies are added to the EOM drive, or as the modulation index at a given frequency is increased. In Fig. S2 the full covariance matrix obtained from the homodyne data is shown on an expanded color scale. A block of the XP covariance matrix is magnified in Fig. S3 to show the structure of the correlations, which show up in the first and fifth off-diagonal elements in this case. The modulation index for each frequency is 0.18. Note that it is comparatively easy to observe the expected structure in the covariance matrix while not being able to reproduce the nullifiers. While the EOM will always move light between frequencies with the chosen frequency shifts, things like noise in the lock circuit can add noise and quickly degrade the nullifier variances to the 2-mode squeezing levels measured with the EOM on, or worse.

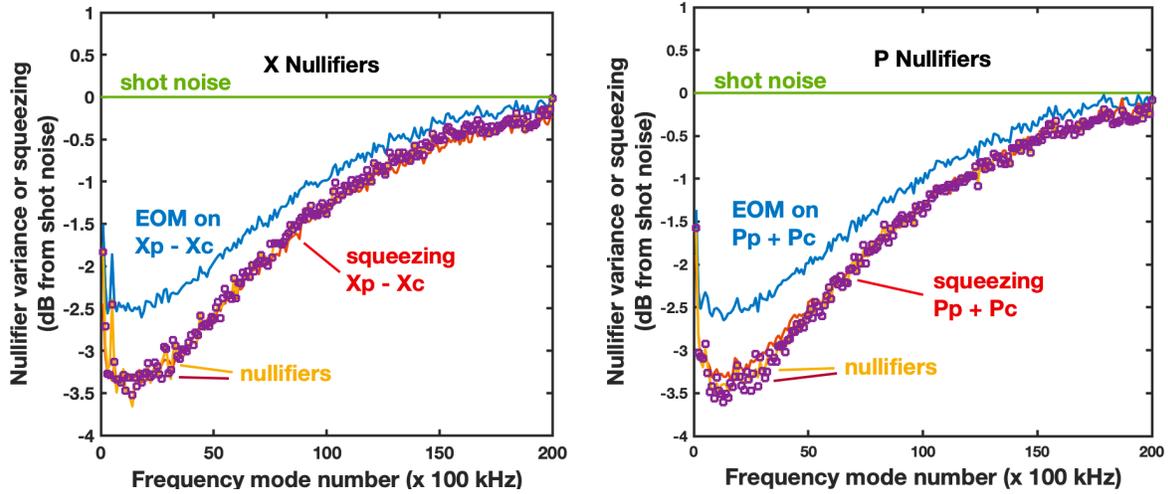

**Fig. S1.** Measurements of 2-mode squeezing and nullifier variances as a function of mode number for a 2-D cluster state with modulation frequencies of 100 kHz and 500 kHz. The bin centers are separated by 100 kHz on the horizontal axis, spanning 20 MHz in frequency. The red curves are the 2-mode squeezing variances with the EOM off ($X_p - X_c$ on the left and $P_p + P_c$ on the right) for the individual frequency modes. The blue curves represent the 2-mode squeezing variances for these modes with the EOM on with a modulation index of 0.18 at each frequency. The points and the gold lines indicate the appropriate X- and P-nullifier variances calculated for these modes; the lines are from a simple covariance calculation and the points are from a numerical lock-in detection calculation of the XP correlations.

The nullifiers for a 4-dimensional cluster state driven with sine wave frequencies of 33 kHz, 99 kHz, 297 kHz, and 891 kHz, and a modulation index at each frequency of $m = 0.18$, are shown in Fig. S4. The data is averaged over 12 traces of 10 ms each. Longer measurements may be required to ensure that the graph state remains well-behaved as the squeezing is increased, as discussed in Ref. (*1, 32*). The width of the frequency bins used here play a large part in determining the amount of signal that is averaged over, and the amount of averaging required for reasonable statistics. A portion of the covariance matrix displaying the correlation structure of the state is shown in Fig. S5.

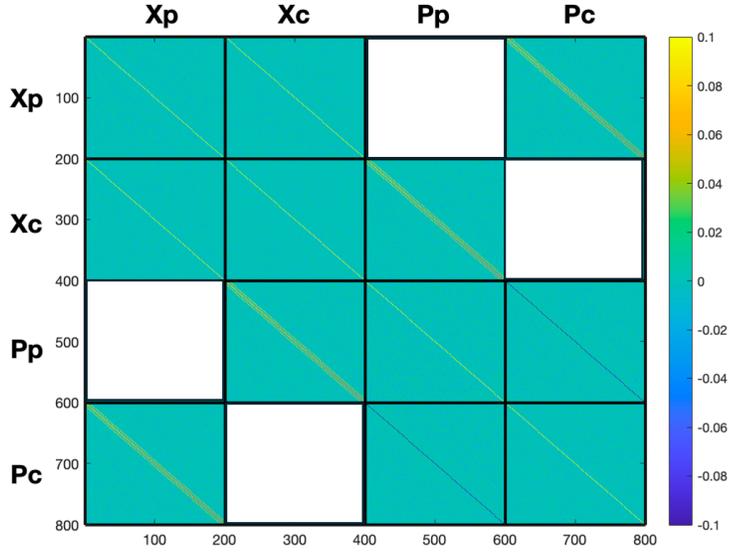

**Fig. S2.** Covariance matrix for a 2-D cluster state. Full covariance matrix on an expanded z-axis (color bar) scale, to show the structure for 100 kHz + 500 kHz, 2-D modulation with m = 0.18 for each frequency. The data is averaged over 12 traces of 10 ms each. The elements of the same-beam measurement blocks XpPp, XcPc, PpXp, and PcXc, indicated in white, are not measured and assumed to be zero. The axes are labeled by mode number.

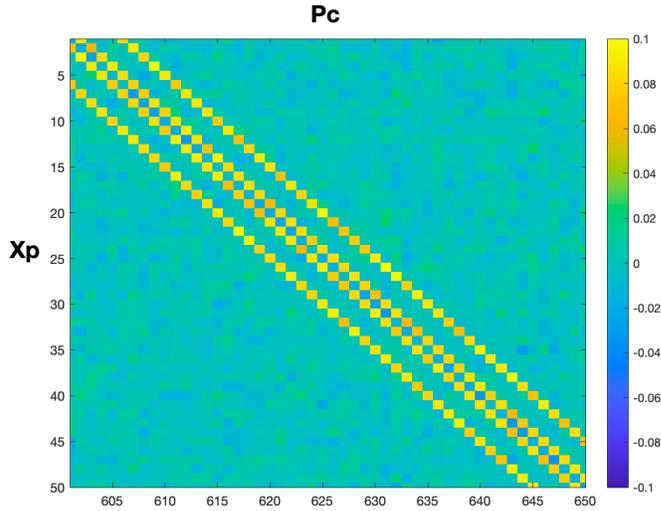

**Fig. S3.** Magnified z-axis view of a portion of the $X_pP_c$ sector of the 2D covariance matrix. This is a part of the covariance matrix shown in Fig. S2, showing the detailed correlation structure. The axes are labeled by mode number.

The fluctuations in the squeezing variances for the individual modes (the 2-mode squeezing) are dominated by the statistical uncertainties in the measurements. The fluctuations in the nullifier variances in Fig. 2, for example, are somewhat larger than those seen in the 2-mode squeezing variances. Partly this is due to the nullifiers drawing on data from neighboring modes that do not have the same squeezing strength. This increases with higher-dimensional graph states, as the nullifiers then include data from frequency bins that are even further away. The increased fluctuations can also be due to the mixing-in of the wrong-quadrature data if the modulation phase(s) are not uniform and well-aligned with the data collection, as the nullifiers draw upon the phase-sensitive XP correlation measurements. Finally, large spikes appear in the nullifiers at the frequency multiples (3, 9, and 27) of the lowest mode index, where the excess noise in this bin gets mixed into the calculated nullifier variances.

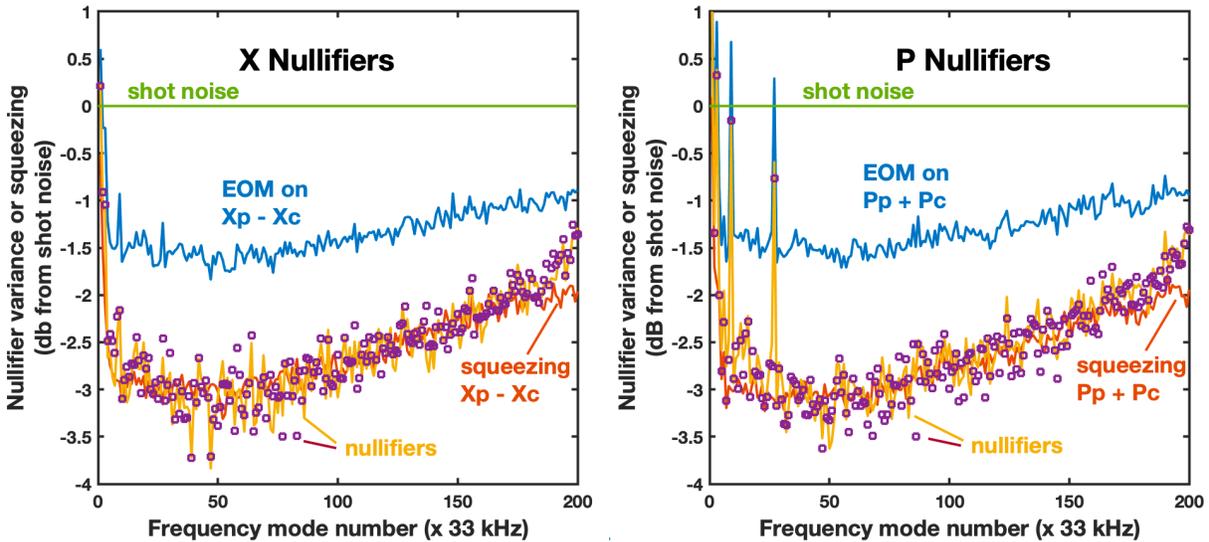

**Fig. S4.** Measurements of 2-mode squeezing and nullifier variances as a function of mode number for a 4-D cluster state. The state is constructed with modulation frequencies of 33 kHz + 99 kHz + 297 kHz + 891 kHz and a modulation index of 0.18 at each frequency. The bin centers are separated by 33 kHz on the horizontal axis, spanning 6.6 MHz in frequency. The data is averaged over 12 traces of 10 ms each. The red curves are the 2-mode squeezing variances with the EOM off ($X_p - X_c$ on the left and $P_p + P_c$ on the right) for the particular frequency modes. The blue curves represent the 2-mode squeezing variances for these modes with the EOM on. The points and the gold lines indicate the appropriate X- and P-nullifier variances calculated for these modes; the lines are from a simple covariance calculation and the points are from a numerical lock-in detection calculation of the XP correlations.

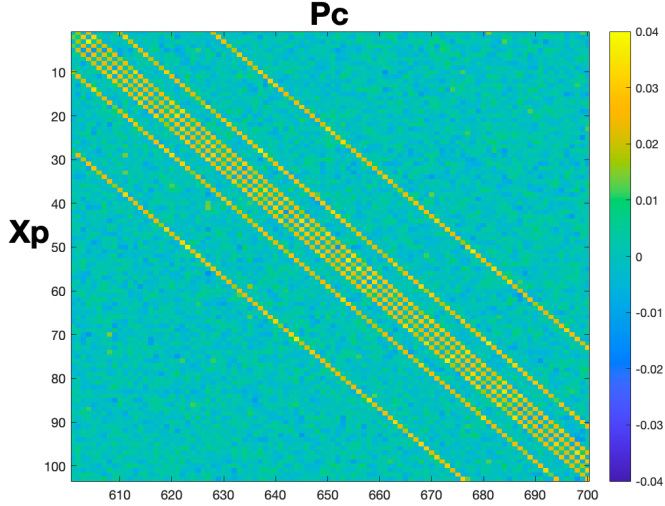

**Fig. S5.** Magnified z-axis (color scale) view of a portion of the XpPc sector of the covariance matrix for the same 4D cluster state for which the nullifiers are shown in Fig. S4. The axes are labeled by mode number.

Nullifier calculations
We wish to calculate the variance of the nullifiers (the variance-based entanglement witnesses) for the resulting state in order to demonstrate the entanglement of the constructed graph state and evaluate how well the quantum properties are preserved. We follow the treatments given in Refs. (*1, 32*) to calculate the nullifiers for a given modulation on the EOM(s).

Formally, a nullifier is an operator for which a particular graph state is an eigenstate with eigenvalue zero. Nullifiers are specific and determined by the state itself, but they are not unique, as linear combinations of nullifiers are also nullifiers. There is one nullifier per mode but note that here the "mode" references a 2-mode squeezed state, so that we have as many nullifiers as we have X and P frequency bins, or twice the number of frequency bins considered. (The "X nullifiers" defined for the probe modes, however, are not independent of those defined for the conjugate frequency modes, as they would correspond to, for instance Xp – Xc and Xc – Xp for the 2-mode squeezed state.)

Practically, a nullifier is a particular linear combination of quadrature operators which exhibits squeezing. What we refer to as the "EPR nullifiers" for this system are the variances of the two-mode squeezing measures for each frequency bin (index *i*); essentially the amplitude-difference and phase-sum quadrature measures: $\left(\hat{X}_p^{(i)} - \hat{X}_c^{(i)}\right)$ and $\left(\hat{P}_p^{(i)} + \hat{P}_c^{(i)}\right)$. Squeezing in these variables is a result of just the two-mode squeezing (TMS) operation and does not mix frequencies. If we do something to further entangle the modes (like introduce an EOM) that would change the nullifiers because that would change the state and its entanglement structure. More formally we can write for discrete and continuous-variable systems:

$$(\vec{\hat{P}} - \mathbf{A}\vec{\hat{X}})|\Psi\rangle = \vec{0}|\Psi\rangle$$
$$(\vec{\hat{P}} - \mathbf{V}\vec{\hat{X}})|\Psi\rangle = \vec{0}|\Psi\rangle \qquad \text{(S2)}$$

The **A** matrix is the adjacency matrix for a discrete graph state, and it is analogous to the complex weighted adjacency matrix **Z** = **V** + i**U** for Gaussian states. Here **V** is the continuous-variable adjacency matrix and **U** is the error matrix. In the limit of perfect squeezing **U** → 0. The operators P – ZX are the nullifiers, but because the matrix **Z** is in general complex, only Gaussian states where **Z** = **V** and **U** = 0 (i.e., **Z** is Hermitian) represent cluster states with real, measurable eigenvalues. Measurement-based quantum computing is still possible, however, when **U** ≠ 0 by using the approximate nullifiers P – **V**X (*1*). These nullifiers satisfy:

$$\text{cov}[P-\mathbf{V}X] = (1/2)\mathbf{U}, \qquad \text{(S3)}$$

where cov[ ] indicates the covariance. Further restrictions on the **V** matrix, to eliminate hidden entanglement that would complicate using such states for measurement-based computing, are discussed in (*30*) and below in the section on GLU transformations. For a proper cluster state **U** → 0 for large squeezing, and in general the nullifier variances will reproduce the 2-mode squeezing levels that we start with. Thus, we need to construct the covariance matrix for the appropriate state to evaluate the nullifiers.

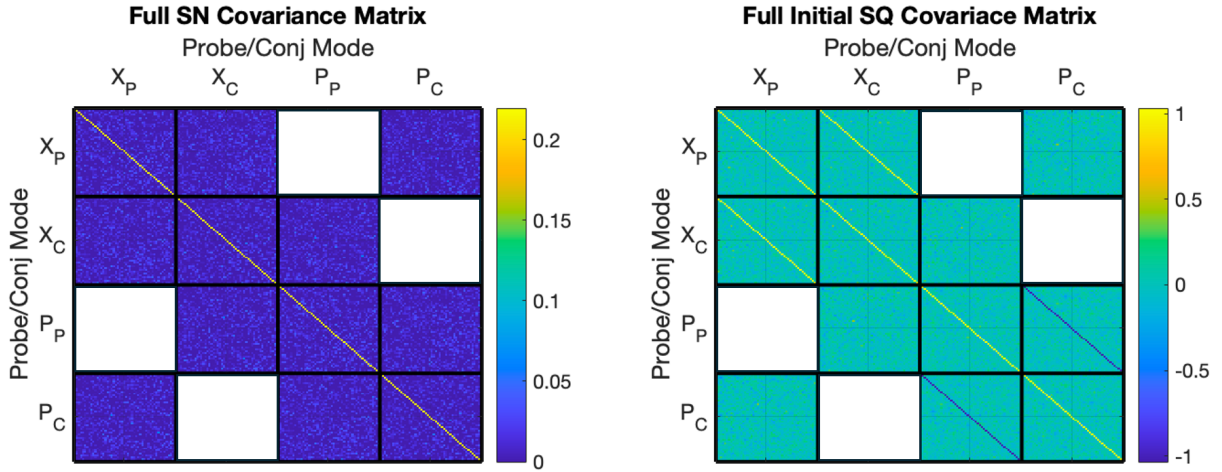

**Fig. S6.** Measured full covariance matrix. A full covariance matrix is shown for the vacuum state (local oscillator shot noise on the homodyne detectors, left) and for a two-mode squeezed state (right). Color scales are enhanced for visibility. White blocks indicate unmeasured correlations.

Theoretically, we generate the covariance matrix Σ by starting with the vacuum state and applying the symplectic transformation matrix for each operation. Σ is proportional to the identity matrix for a vacuum state (autocorrelations on the diagonal and no off-diagonal correlations). We can then begin with the covariance matrix generated by the two-mode squeezing (TMS) operation: $\Sigma = S_{TMS} S^T_{TMS}$, shown in Fig. S6, and then chain the various successive operations to this.

We apply the EOM transformation matrix to obtain the new covariance matrix:

$$\Sigma = (1/2) S_{eom} S_{TMS} S_{vac} S^T_{vac} S^T_{TMS} S^T_{eom} \tag{S4}$$

For an EOM in the conjugate beam, defining matrices X1 and Y1:

$$\begin{aligned} X1_{i,j} &= J_0(m), & Y1_{i,j} &= 0 & \text{for } i=j \\ X1_{i,j} &= 0, & Y1_{i,j} &= J_1(m) & \text{for } i = j \pm 1 \\ X1_{i,j} &= 0, & Y1_{i,j} &= 0 & \text{for all other } i,j \end{aligned} \tag{S5}$$

$$A1 = \begin{pmatrix} 1 & 0 \\ 0 & X1 \end{pmatrix} \quad B1 = \begin{pmatrix} 0 & 0 \\ 0 & Y1 \end{pmatrix} \tag{S6}$$

and $S_{eom} = \begin{pmatrix} A1 & B1 \\ -B1 & A1 \end{pmatrix}$, (S7)

This expression includes only the first sidebands, as we assume a small enough modulation index for these to be the only connections that matter (1). The symplectic matrix, $S_{eom}$ defined above, for a single-frequency, single-EOM (in the conjugate beam) with a modulation index of $m = 0.36$ is shown in Fig. S7. The weights of the terms in the indicated blocks are given. Here the weights of the X's and P's of the conjugate modes are proportional to Bessel functions, as in Eq. (S1). Those on the diagonal are reduced to $J_0(m) = 0.967$, and the pairs of sidebands that show up have weights that are proportional to $J_1(m) = 0.178$. (Higher-order sidebands could be included but are very small at the modulation indices used here.) The corresponding covariance matrix is shown in Fig. S8. Note that this covariance matrix is highly symmetric, even though the EOM was placed in a single beam (the conjugate). Given the symmetry present, one could expect that the same extra correlations would appear if the EOM were placed in the probe beam, and this is in fact the case.

The matrix that contains the nullifier vectors for these modes is constructed by first taking $N_{TMS} = I * S^{-1}_{TMS}$, where I is the identity matrix. The form of $S_{TMS}$, the symplectic transformation for a pair of modes for the two-mode squeezing operation, is:

$$S_{TMS} = \begin{pmatrix} \cosh(2r) & \sinh(2r) & & 0 \\ \sinh(2r) & \cosh(2r) & & \\ & & \cosh(2r) & -\sinh(2r) \\ 0 & & -\sinh(2r) & \cosh(2r) \end{pmatrix}, \tag{S8}$$

where r is the squeezing parameter. The nullifier matrix after the application of an EOM is then given by:

$$N_{eom} = N_{TMS} * S^{-1}_{eom}. \tag{S9}$$

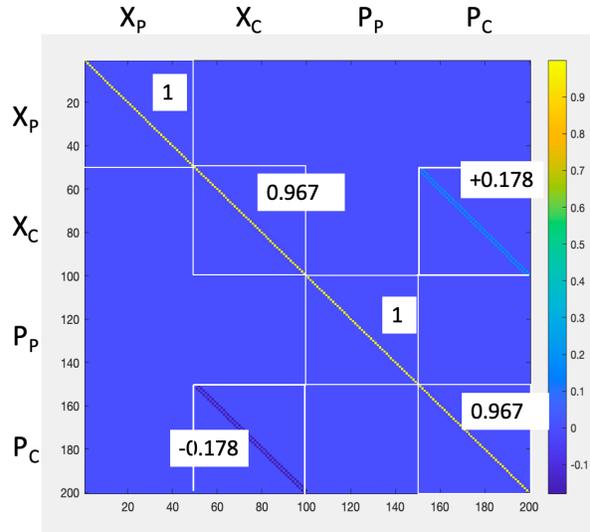

**Fig. S7.** An example of the $S_{eom}$ symplectic matrix for an EOM in the conjugate beam. The mode ordering is in blocks: Xp, Xc, Pp, Pc. The values indicated represent Bessel functions $J_0(m)$ and $J_1(m)$ for a modulation index $m = 0.36$. (This $m$ corresponds to 60 V in our case.) The diagonal elements in the diagonal blocks are equal to either 1 or $J_0(m)$, and the values shown in the XP blocks equal $\pm J_1(m)$.

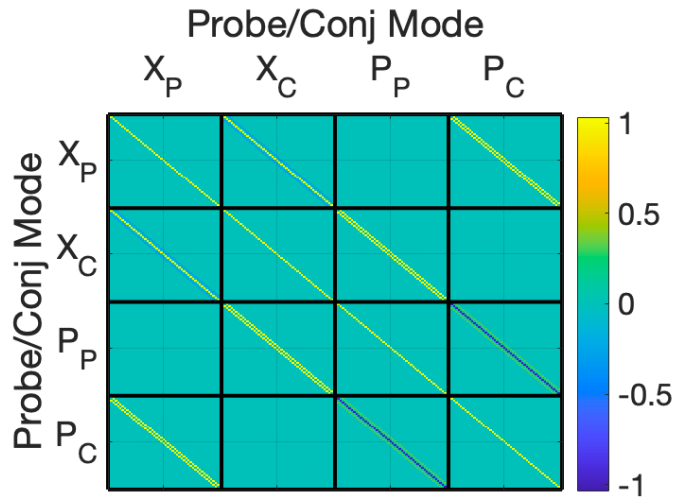

**Fig. S8.** The expected covariance matrix for the case of an EOM in the conjugate beam. The EOM is driven by a single frequency with a modulation index $m = 0.36$. (The symmetry evident in this covariance matrix indicates that it does not matter if we place the EOM into the probe beam or the conjugate beam.) The color scale is enhanced for visibility.

The rows of this matrix contain the weighted distribution of the correlations amongst the modes. There is one nullifier vector for each frequency mode, and nullifiers are defined for both the X and the P quadratures. Thus, for M frequency modes, we will calculate 4M "nullifier variances", as each frequency has corresponding X and P nullifiers. This double-counts the modes, as it would, for instance count $(X^{(i)}_{probe} - X^{(i)}_{conj})$ and $(X^{(i)}_{conj} - X^{(i)}_{probe})$ separately, as mentioned above. Thus, there are 2M independent nullifiers for the M frequency modes. Linear combinations of such nullifiers are also squeezed. Since the squeezing varies over the spectrum of our frequency modes, and here the EOM mixes only a small fraction of other modes into the original frequency modes, we associate the nullifiers with the mode labels of the original two-mode-squeezing frequency basis.

To calculate the nullifier variances we use the row vector of the nullifier matrix for each quadrature mode. For the $i^{th}$ mode we take the row vector made from the $i^{th}$ row of $N_{eom}$, denoted $[N_{i,*}]$, and calculate the matrix,

$$[N_{i,*}]^T * [N_{i,*}], \qquad (S10)$$

and then multiply this matrix by the covariance matrix $\Sigma$, and then sum over all the matrix terms. Equivalently, we can calculate $[N_{i,*}]^T * \Sigma * [N_{i,*}]$, resulting in the variance for the chosen frequency mode and quadrature.

For the example of a single EOM driven with a modulation index of 0.36 we plot the nullifier matrix in Fig. S9 and highlight the calculation of the X nullifier for the 18th mode. The 18th row of the nullifier matrix is plotted there as well. In Fig. S10 we plot the matrix $[N_{18,*}]^T * [N_{18,*}]$, and indicate the corresponding Bessel function values of the terms.

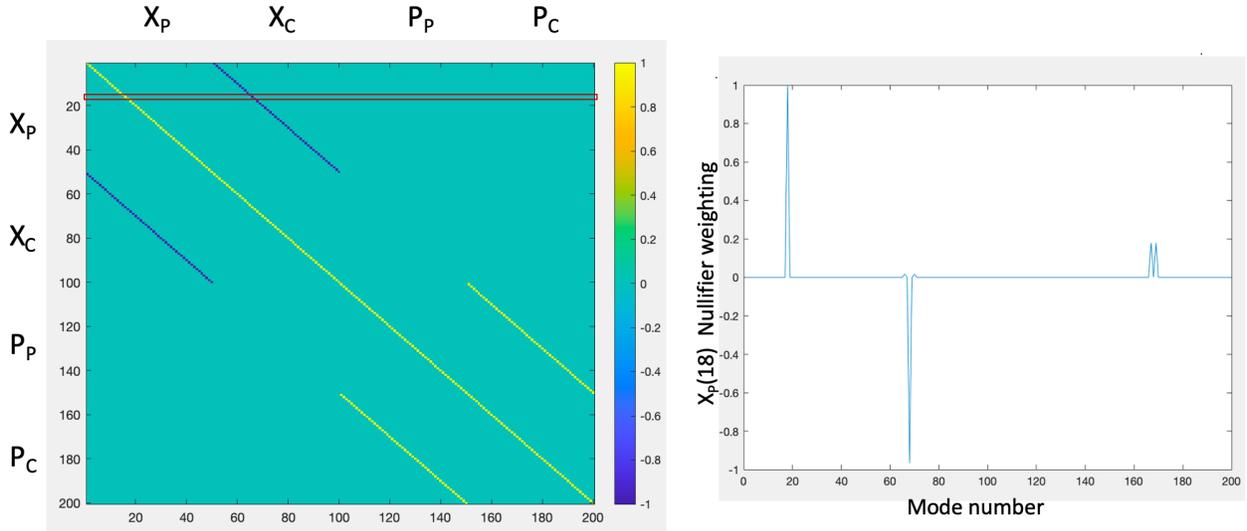

**Fig. S9.** Example nullifier matrix $N_{eom} = N_{TMS} * S^{-1}_{eom}$ showing the structure of the nullifiers. $N_{eom}$ with row 18 highlighted (left) and the values in row 18 plotted (right). (The color scale is expanded in the left panel.)

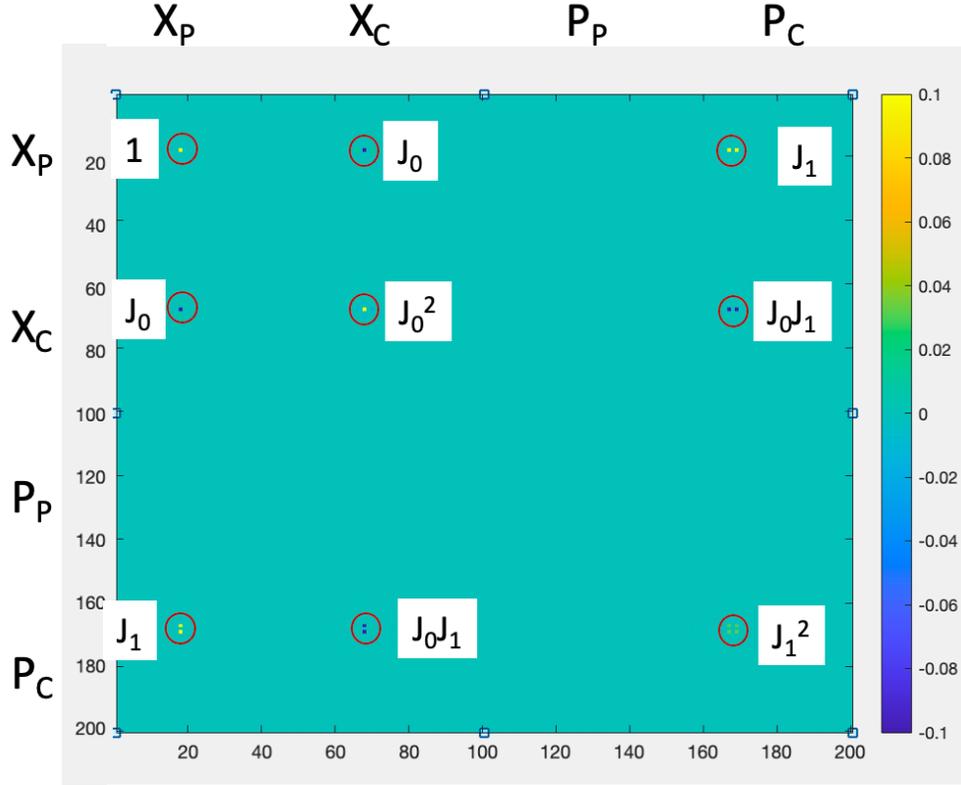

**Fig. S10.** The matrix $[N_{18,*}]^T * [N_{18,*}]$, which sets out the X nullifier terms for the 18th mode. The color scale is expanded in the figure. The non-zero terms are highlighted in the small circles and the terms are proportional to the indicated Bessel function products evaluated at the modulation index of 0.36. The color scale is enhanced for visibility.

Summing over the terms in Fig. S10 gives the nullifier formula; the non-zero terms are highlighted in the small circles in the figure and the terms are proportional to the indicated Bessel function products evaluated at the modulation index. Adding together the weighted terms of the covariance matrix by multiplying this matrix by the covariance matrix gives the appropriate nullifier variance. The directly calculated nullifiers show a large variation due to the frequency-dependence of the electronic noise in the detection system. We can remove this variation by subtracting the electronic noise from the detectors or by normalizing to the shot noise, which is what is done in Figs. 2 and 4 of the main text and Figs. S1 and S4.

Alternate approach to nullifier derivation
A simpler way to see where the nullifiers come from for a single modulation frequency is to examine the 2-mode squeezing X and P "EPR nullifiers," and substitute the appropriate single-sideband-mixing approximation for the EOM-modified modes into these formulae to derive the nullifiers. This is a direct calculation for single-frequency modulation but becomes cumbersome for more complex situations. We start with the expressions for 2-mode squeezing nullifiers:

Xp – Xc
Pp + Pc                                                                                              (S11)

The action of the EOM is to "rotate" light at the sideband frequencies from the +P quadrature into the +X quadrature, and from the +X quadrature into the –P quadrature (defining the P as the vertical or y-axis and the X as the horizontal or x-axis, and rotation clockwise). Thus, we substitute,

$$X'c(i) \approx J_0(m) Xc(i) - J_1(m) Pc(i-1) - J_1(m) Pc(i+1)$$
$$P'c(i) \approx J_0(m) Pc(i) + J_1(m) Xc(i-1) + J_1(m) Xc(i+1) , \qquad (S12)$$

where $i$ is the mode index, $J_0$ and $J_1$ are Bessel functions, and $m$ is the modulation index, which we will suppress in what follows.

The nullifier variances then become:

$$\begin{aligned}
\langle [Xp - X'c]^2 \rangle &= \langle [Xp(i) - (J_0 Xc(i) - J_1 Pc(i-1) - J_1 Pc(i+1))]^2 \rangle \\
&= \langle Xp(i) Xp(i) \rangle - 2 \langle Xp(i) (J_0 Xc(i) - J_1 Pc(i-1) - J_1 Pc(i+1)) \rangle \\
&\quad + \langle (J_0 Xc(i) - J_1 (Pc(i-1) + Pc(i+1)))^2 \rangle \\[4pt]
&= \langle Xp(i) Xp(i) \rangle - J_0 \langle Xp(i) Xc(i) \rangle + J_1 \langle Xp(i) Pc(i-1) \rangle + J_1 \langle Xp(i) Pc(i+1) \rangle \\
&\quad - J_0 \langle Xc(i) Xp(i) \rangle + J_1 \langle Xp(i) Pc(i-1) \rangle + J_1 \langle Xp(i) Pc(i+1) \rangle \\
&\quad + J_0^2 \langle Xc(i) Xc(i) \rangle - 2 J_0 J_1 \langle Xc(i)(Pc(i-1) + Pc(i+1)) \rangle \\
&\quad + \langle (- J_1 Pc(i-1) - J_1 Pc(i+1))^2 \rangle \\[4pt]
&= \langle Xp(i) Xp(i) \rangle - J_0 \langle Xp(i) Xc(i) \rangle + J_1 \langle Xp(i) Pc(i-1) \rangle + J_1 \langle Xp(i) Pc(i+1) \rangle \\
&\quad - J_0 \langle Xc(i) Xp(i) \rangle + J_0^2 \langle Xc(i) Xc(i) \rangle - J_0 J_1 \langle Xc(i)(Pc(i-1) + Pc(i+1)) \rangle \\
&\quad + J_1 \langle Xp(i) Pc(i-1) \rangle - J_0 J_1 \langle Xc(i) Pc(i-1) \rangle + J_1^2 \langle Pc(i-1)(Pc(i-1) + Pc(i+1)) \rangle \\
&\quad + J_1 \langle Xp(i) Pc(i+1) \rangle - J_0 J_1 \langle Xc(i)(Pc(i+1) \rangle \\
&\quad + J_1^2 \langle Pc(i+1)(Pc(i-1) + Pc(i+1)) \rangle \qquad (S13)
\end{aligned}$$

and

$$\begin{aligned}
\langle [Pp + P'c]^2 \rangle &= \langle [Pp(i) + (J_0 Pc(i) + J_1 Xc(i+1) + J_1 Xc(i-1))]^2 \rangle \\[4pt]
&= \langle Pp(i) Pp(i) \rangle + 2 \langle Pp(i)(J_0 Pc(i) + J_1 Xc(i+1) + J_1 Xc(i-1)) \rangle \\
&\quad + \langle (J_0 Pc(i) + J_1 Xc(i+1) + J_1 Xc(i-1))^2 \rangle \\[4pt]
&= \langle Pp(i) Pp(i) \rangle + J_0 \langle Pp(i) Pc(i) \rangle + J_1 \langle Pp(i) Xc(i-1) \rangle + J_1 \langle Pp(i) J_1 Xc(i+1) \rangle \\
&\quad + J_0 \langle Pc(i) Pp(i) \rangle + J_1 \langle Pp(i) Xc(i-1) \rangle + J_1 \langle Pp(i) Xc(i+1) \rangle \\
&\quad + J_0^2 \langle Pc(i) Pc(i) \rangle + 2 J_0 J_1 \langle Pc(i)(Xc(i+1) + Xc(i-1)) \rangle \\
&\quad + \langle (J_1 Xc(i-1) + J_1 Xc(i+1))^2 \rangle \\[4pt]
&= \langle Pp(i) Pp(i) \rangle + J_0 \langle Pp(i) Pc(i) \rangle + J_1 \langle Pp(i) Xc(i-1) \rangle + J_1 \langle Pp(i) J_1 Xc(i+1) \rangle \\
&\quad + J_0 \langle Pc(i) Pp(i) \rangle + J_0^2 \langle Pc(i) Pc(i) \rangle + J_0 J_1 \langle Pc(i)(Xc(i-1) + Xc(i+1)) \rangle \\
&\quad + J_1 \langle Xc(i-1) Pp(i) \rangle + J_0 J_1 \langle Pc(i) Xc(i-1) \rangle + J_1^2 \langle Xc(i-1)(Xc(i-1) + Xc(i+1)) \rangle \\
&\quad + J_1 \langle Xc(i+1) Pp(i) \rangle + J_0 J_1 \langle Pc(i) Xc(i+1) \rangle \\
&\quad + J_1^2 \langle Xc(i+1)(Xc(i-1) + Xc(i+1)) \rangle . \qquad (S14)
\end{aligned}$$

With this arrangement of terms the equivalence of Eq. (S13) to the terms summed in Fig. S10 becomes apparent.

Electronic noise subtraction/normalization to shot noise
The 2-mode squeezing noise measurements are obviously a noise power plotted as a function of frequency. To correct such measurements for the electronic noise background one will often measure an intensity-difference spectrum directly on an RF spectrum analyzer, and then measure the electronic noise background with no signals on the detectors and subtract the electronic noise spectrum from the 2-mode squeezing spectrum. Here it is somewhat less obvious that the nullifier variances are equivalent to power spectra but, ordered as they are, they become identical to the 2-mode squeezing spectra in the limit of small modulation. As long as the electronic noise does not change much over the frequency span of the sidebands that are mixed into the nullifier expressions, this subtraction can be performed. In our measurements there was about 6 dB of clearance between the electronic noise and the shot noise level, resulting in the electronic noise being only about 3 dB below the squeezing signals, making the corrections useful.

An essentially equivalent means of normalizing the signals is to divide the signals by the shot noise signal, which ought to be flat as a function of frequency. To within the noise on the measurements these two methods produce equivalent results, although with the electronic noise subtraction a slight upward slope with frequency remains on all the signals. We present the signals normalized to shot noise.

Calculation of Gaussian local unitary transformations
The graph states that are directly generated by sinusoidal phase modulation with an EOM are not generally in the form of cluster states, but some are transformable into true cluster states by implementing a series of Gaussian local unitary (GLU) operations on the qumodes (*1*). A procedure to determine these operations, when possible, is given in (*30*). For instance, by giving a $-\pi/2$ rotation to every even probe mode and every odd conjugate mode, the 1-D state that is directly generated by single-frequency modulation with an EOM is transformed into the form of a cluster state.

The same GLU transformation that changes the 1-D graph into the form of a cluster state also works on higher-dimensional states, with the important provision that the higher frequencies are odd multiples of the lower frequencies. Thus, the states shown here with, say, 100 kHz, 300 kHz and 900 kHz driving frequencies on the EOM are directly transformable into 3-D cluster states. Interestingly, states generated with even-multiple frequencies are not transformable in this way. In this case we generate graph states that contain hidden entanglement, as discussed in (*30*). This will be shown explicitly in a future publication.

Cluster state structure
Finally, we can plot the adjacency matrix of Eq. (S2) in a plot that shows the entanglement connections and their weights in Figs. S11 - S14 for the 1-D through 4-D experiments above. Extraneous connections appear in the experimental adjacency matrices, due primarily to the effects of measurement noise. With better measurement statistics the desired interconnections become more uniform and the extraneous connections fade. Note that we do not show the probe-conjugate connections (between the nodes in each of the probe-conjugate node-pairs in the figures) that the 4WM process generates, as they are much stronger than the EOM-generated entanglement connections and would dominate. In dimensions higher than one there are traceback connections (for instance, from one horizontal row to the next in Fig. S12) that alter the cluster state structure, and these edge modes would need to be removed to make a proper lattice for computing, as discussed in Ref. (*1*). In Fig. S13, for the 3-D cluster state, we include a plot of the theoretical adjacency graph with an amount of noise approximately equal to the experimental noise added to the theoretical covariance matrix.

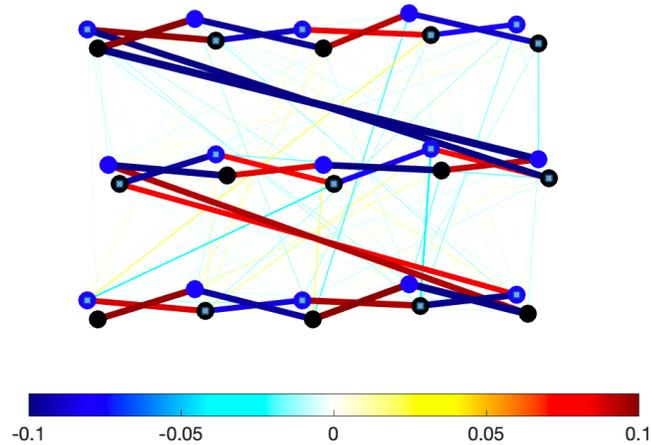

**Fig. S11**. This figure shows a 1 x 15 section of the adjacency connections for a 1D experiment with 100 kHz modulation and a modulation index of 0.18 and averaged over 12 runs of 10 ms. The adjacent node-pairs are the probe (blue dots) and conjugate (black dots) modes at the same sideband frequency and are entangled by the 2-mode squeezing generated in the 4-wave mixing. The colored bars indicate the strength of the mixing or entanglement induced by the EOM. The probe-conjugate entanglement at the same sideband frequency is not shown, as it would be much stronger than the displayed links. Extraneous connections between the non-neighboring mode-pairs are due to measurement noise. The 1D chain of nodes is folded into sections of 5 node pairs to compare with the 2D cluster of Fig. S12.

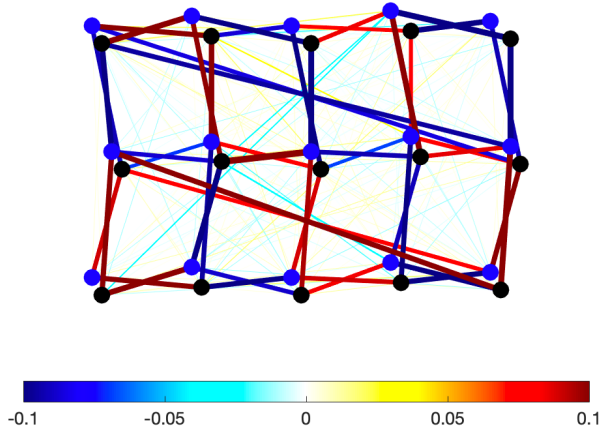

**Fig. S12.** This figure shows a 3 x 5 section of the adjacency connections for a 2D experiment with modulation frequencies of 100 kHz and 500 kHz, a modulation index of 0.18 for each frequency, and averaged over 12 runs of 10 ms. The colored bars indicate the strength of the mixing or entanglement induced by the EOM. The adjacent node-pairs are the probe (blue dots) and conjugate (black dots) modes at the same sideband frequency. Extraneous connections between the non-neighboring mode-pairs on the lattice are due to measurement noise.

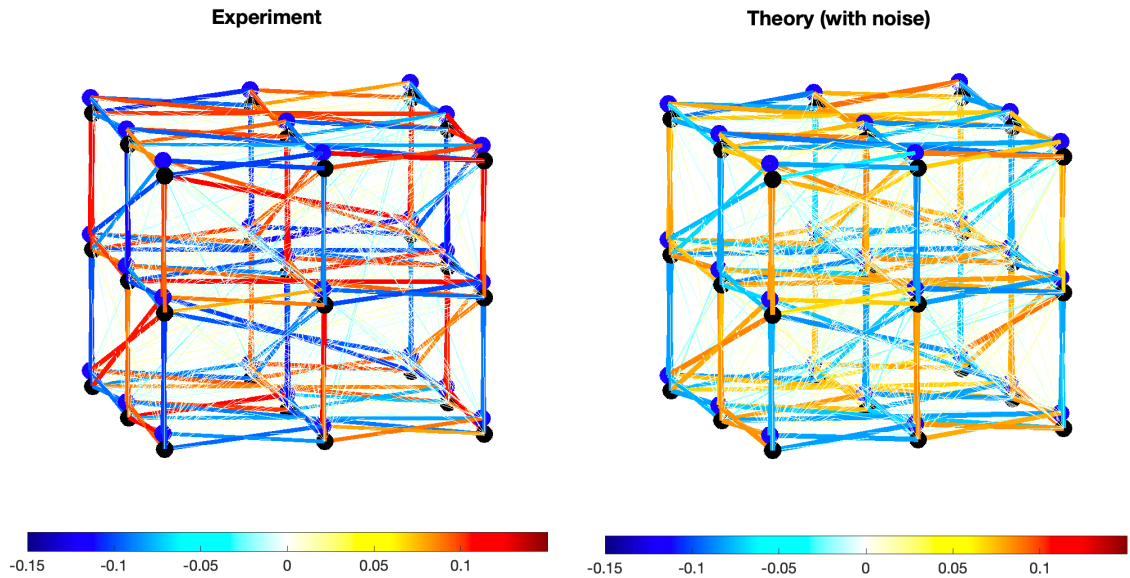

**Fig. S13.** A graphical representation of the entanglement (adjacency matrix) between the nodes of the experimental 3-D cluster state of Figs. 3 and 4 in the main text. The plots show a 3 x 3 x 3 portion of (left) the experimental 3-D cluster state. The adjacent node-pairs are the probe (blue dots) and conjugate (black dots) modes at the same sideband frequency and are entangled by the 2-mode squeezing generated in the 4-wave mixing. The EOM is driven at 100 kHz, 300 kHz, and 900 kHz, each with a modulation index of 0.18, averaged over 24 runs of 10 ms each. A theoretical prediction for this case is shown (right) with nose added to the covariance matrix at the same level as found in the experiment. The color bars indicate the strength of the mixing or entanglement induced by the EOM.

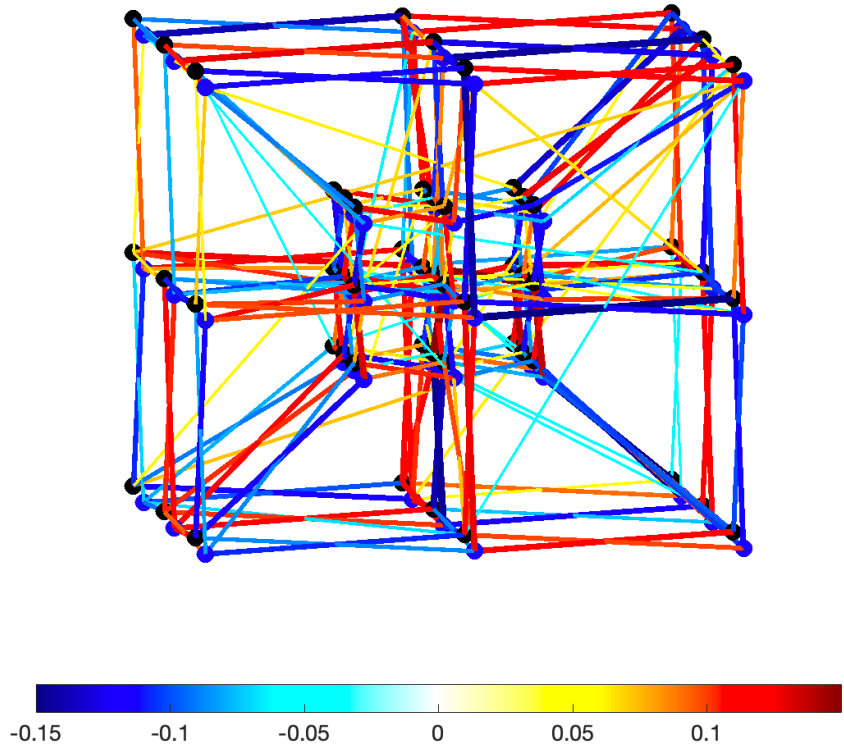

**Fig. S14.** A graphical representation of the entanglement (adjacency matrix) between the nodes of the experimental 4-D cluster state of Figs. S4 and S5. The plots show a 3 x 3 x 3 x 3 portion of the experimental 4-D cluster state (represented as a tesseract), whose nullifiers are plotted in Fig. S4. The adjacent node-pairs are the probe (blue dots) and conjugate (black dots) modes at the same sideband frequency and are entangled by the 2-mode squeezing generated in the 4-wave mixing. The connections are induced between the frequency modes by driving the EOM at 33 kHz, 99 kHz, 297 kHz, and 891 kHz, each with a modulation index of 0.18. Data from 12 runs of 10 ms each are averaged. The colored bars indicate the strength of the mixing or entanglement induced by the EOM. Connections with a strength less than 0.05 are removed from the plot for clarity of the structure. Variations in the connection strengths are due to measurement noise.